# Study on $CePtIn_4$ grown in a platelet-like morphology

Debarchan Das,[1] Marek Daszkiewicz,[1] Daniel Gnida,[1] Alicja Hackemer,[1]
Mirosław Werwiński,[2] Andrzej Szajek,[2] and Dariusz Kaczorowski[1]*

[1]*Institute of Low Temperature and Structure Research, Polish Academy of Sciences, P. O. Box 1410, 50-590 Wrocław 2, Poland*

[2]*Institute of Molecular Physics, Polish Academy of Sciences, Mariana Smoluchowskiego 17, 60-179 Poznań, Poland*

We present the results of our comprehensive structural investigation performed on single crystals of $CePtIn_4$ grown from In flux in a platelet-like morphology. They were found to crystalize in an orthorhombic structure having space group *Cmcm* (no. 63) with the lattice parameters $a$ = 4.52175 (15) Å, $b$ = 16.7777 (5) Å, and $c$ = 7.3903 (2) Å. The unit cell volume of such crystals is slightly larger than that of previously reported needle-shaped single crystals of $CePtIn_4$. This feature seems to influence the low-temperature physical properties of the compound, such as enhanced ordering temperature ($T_N$ = 2.30(2) K versus 1.65(1) K, observed before) along with a sharp drop in the electrical resistivity at $T_N$ (that was absent for the needle-like $CePtIn_4$). In addition to the experimental data, we present the results of our *ab-initio* calculations of the electronic structure of $CePtIn_4$ with the platelet-like morphology.

*Corresponding author: d.kaczorowski@intibs.pl

## 1. Introduction

For past few decades cerium-bearing compounds have attracted significant research interest due to their diverse intriguing electronic properties, such as heavy-fermion behavior, spin and valence fluctuations, non-Fermi liquid features, complex magnetic ordering, superconductivity, quantum criticality, etc. [1-3]. In recent years, extensive studies were performed on ternary cerium indides, like heavy-fermion superconductors $CeCoIn_5$ [4] and $Ce_2PdIn_8$ [5,6,7], with paramagnetic ground states, heavy-fermion systems $Ce_3PtIn_{11}$ [8-10] and $Ce_3PdIn_{11}$ [11,12], in which superconductivity coexists with long-range antiferromagnetic ordering, as well as non-superconducting antiferromagnetic or ferromagnetic Kondo lattices, such as $Ce_6Pd_{12}In_5$ [13] and $Ce_{11}Pd_4In_9$ [14], respectively.

Most recently, we reported on the formation of a novel phase in the ternary Ce-Pt-In system, namely $CePtIn_4$ [15]. Sizable needle-like single crystals of this compound were grown from In flux. They were found to crystallize with an orthorhombic crystal structure of the $YNiAl_4$-type. Measurements of their electrical transport and thermodynamic properties revealed metallic conductivity, Curie-Weiss paramagnetism due to $Ce^{3+}$ ions with well localized 4*f*-electrons, and antiferromagnetic order below $T_N$ = 1.65(1) K.

Continuing our search for novel Ce-Pt-In ternaries, focused at hypothetical $CePtIn_5$, we obtained single crystals of the $CePtIn_4$ phase, yet with platelet-like morphology, distinctly different from that observed earlier [15]. Here, we report the results of our detailed crystallographic study of these new crystals (platelets, hereafter labeled PL-type) and compare them with those of the crystals examined before (needles, labeled NE-type). We also describe the low-temperature physical properties of the PL-type crystals, which appeared to differ somewhat from those of the NE-type ones. In addition, in order to give an insight into the magnetic properties of $CePtIn_4$ from the perspective of electronic structure, we carried out the ab-initio calculations based on the density functional theory (DFT) [16].

## 2. Methods

Single crystals of $CePtIn_4$ were synthesized using In flux method. The constituent elements (Ce-3N, Pt-4N and In-5N) were taken in the molar ratio of Ce:Pt:In = 1:1:27 and placed in an alumina crucible, which was then sealed in a quartz ampule under 0.3 atm. of argon. The ampule was first heated in a resistance furnace up to 1200 °C, kept at this temperature for 48 hours, and then cooled to 750 °C at a rate of 150 °C/hour, maintained at this temperature for 10 hours, and finally cooled to 550 °C at a rate of 1 °C/hour. At the latter temperature, In flux was decanted using a centrifuge. The so-obtained single crystals had a form of well-developed platelets with typical dimensions 1 mm×1 mm×0.06 mm. In this context it worth to highlight that for the needle-like crystals, the constituent elements were taken in a slightly different molar ratio and the maximum temperature reached during heat treatment was 1000 °C [15]. In Figure 1, we present a photograph of the as-grown crystals. For comparison, we show in that figure also the crystals investigated in Ref. 15, which have a form of needles.

Chemical composition of the single crystals was examined by energy dispersive x-ray (EDX) analysis using a FEI scanning electron microscope equipped with an EDAX spectrometer. The results indicated that the obtained platelet-like single crystals are homogeneous and have the expected stoichiometry.

The X-ray diffraction (XRD) intensities were collected at 295 K on an Oxford Diffraction four-circle diffractometer equipped with an Atlas CCD detector using graphite-monochromatized Mo$K_\alpha$ radiation ($\lambda$ = 0.71073 Å). The raw data were treated with the CrysAlis Data Reduction Program (version 1.171.38.34a). The Bragg intensities were corrected for Lorentz and polarization effects. The absorption was corrected by numerical methods based on Gaussian integration over a multifaceted crystal model and empirical absorption correction using spherical harmonics, implemented in SCALE3 ABSPACK scaling algorithm. The crystal structure was solved by direct methods and refined by full-matrix least-squares method using SHELXS-2014 and SHELXL-2014 programs, respectively [17]. The structure was refined using anisotropic displacement parameters for each atom. Visualizations of the structure were made using Diamond 3.2k [18] and VESTA[19].

Magnetic measurements were performed in the temperature range 1.8 – 300 K using a Quantum Design MPMS-XL SQUID magnetometer. The electrical resistivity was measured over the temperature range 1.8 – 300 K employing standard ac four-probe technique implemented in a Quantum Design PPMS-9 platform. Specific heat measurements were carried out using relaxation method in the same PPMS.

To carry out the DFT calculations the full-potential local-orbital (FPLO) code [20] was chosen in full potential and fully relativistic approach. In the calculations, the exchange-correlation potential provided by Perdew-Burke-Ernzerhof (PBE) [21] within generalized gradient approximation (GGA) was used. For integration over the Brillouin zone (BZ), a grid of 8000 k-points (20x20x20) with the standard tetrahedron method was used [22]. The criteria used for self-consistency were $10^{-8}$ Ha/cell for energy and $10^{-6}$ for charge density. The computer models of crystal structures were based on the lattice constants and atomic positions obtained from the X-ray diffraction measurements made on $CePtIn_4$ single crystals grown in the platelet-like morphology. In order to consider the antiferromagnetic configurations of the magnetic moments on cerium atoms, three structural models with reduced symmetry were prepared, which reproduced the simplest sequences of magnetic moments in Ce atoms layers along the crystallographic *b*-axis, i.e. (− Pt + + Pt −), (− Pt + − Pt +) and (+ Pt + − Pt −), where + and − denote opposite magnetic moments on Ce atoms (hereafter labeled also as Ce1 and Ce2) and the settings were defined in relation to Pt atoms layers in the crystallographic unit cell of $CePtIn_4$. For Ce 4*f* orbitals, the orbital polarization corrections [23, 24] were included in order to improve the model in respect of the orbital magnetic moments usually underestimated in GGA.

## 3. Results and Discussion

### 3.1. Crystal structure

The experimental details on the structure refinement of the PL-type $CePtIn_4$ crystal along with the main crystallographic data are gathered in Table 1. The atomic coordinates and equivalent isotropic displacement parameters are given in Table 2, while the anisotropic displacement parameters are listed in Table 3. The refined lattice parameters for PL- type crystals are slightly larger than those derived for the NE-type crystals (with the unit cell volume smaller by about 1.6%) [15], and more similar to the values reported recently for Zn-stabilized $CePtIn_4$ (having the unit cell volume smaller by about 0.5% with respect to that of the PL-type crystals) [25]. In the present case, In flux was used for crystal growth, so no occupational substitution of any foreign atom can occur. One should also note that all the anisotropic displacement parameters are slightly smaller for the PL-type crystals than for the NE-type ones. This finding suggests that all the atoms create a well-defined network. The crystal structure of the compound is shown in Figure 2, where also the coordination spheres of Ce and Pt atoms are depicted. Both atoms are basically coordinated by In atoms, with eleven and seven In neighbors, respectively (cf. Figure 2). Within the coordination sphere of the Ce atom one finds also two Pt atoms. They may play a stabilizing role in the structure. Since the unit cell dimensions of the PL-type crystals are slightly larger than those of the NE-type one, also the interatomic distances are somewhat longer in the former modification of $CePtIn_4$ (see Table 4).

### 3.2. Physical properties

For the magnetic studies, several single-crystalline platelets of CePtIn4 were mounted in a planar manner on a sapphire substrate using Apiezon-N grease, with their crystallographic *b* axis aligned along magnetic field direction. The contribution of sample holder to the measured signal was found sizeable, yet almost temperature independent. Figure 3 shows the temperature variation of the inverse magnetic susceptibility, $\chi^{-1}(T)$, after subtracting the sample holder contribution. Above 50K, the experimental data can be well modeled by the Curie-Weiss (CW), $\chi\,(T) = \frac{C}{T-\theta_W}$, where $C$ is the Curie constant related to the effective magnetic moment as $\mu_{\mathrm{eff}} = \sqrt{8C}$, $\theta_W$ stands for the Weiss temperature. The red solid line in Figure 2 represents fitting of CW to the measured data that yielded: $\mu_{\mathrm{eff}}$ = 2.62(1) $\mu_B$, $\theta_W$ = −7.6(9) K. The so-obtained value of $\mu_{\mathrm{eff}}$ is close to the theoretical value of 2.54 $\mu_B$, calculated for a free $Ce^{3+}$ ion within the Russell-Saunders coupling scenario, which in turn suggests that 4*f* electrons in $CePtIn_4$ are fairly well localized. In turn, the negative value of $\theta_W$ indicates antiferromagnetic (AFM) exchange interactions.

Indeed, as can be inferred from the upper inset to Figure 3, the sample exhibits a pronounced maximum in $\chi(T)$ at $T_N$ = 2.30(2) K, characteristic of AFM phase transition. Remarkably, the Neel temperature of the PL-type $CePtIn_4$ is significantly higher than $T_N$ = 1.65(1) K determined before for the NE-type modification of the compound [15] and also higher than $T_N$ = 1.9 K reported for the Zn-containing crystals [25]. In order to confirm the AFM character of the ordered state, isothermal magnetization measurements were carried out at temperatures below and above $T_N$. As displayed in the lower inset to Figure 3, the isothermal magnetization, $M(H)$, taken at 1.8 K with field applied along the crystallographic *b* axis, exhibits a clear deviation from linear dependency at around 11 kOe indicating a metamagnetic-like transition, which is a behavior often observed for antiferromagnets. In strong field, *M*(*H*) tends to saturate towards a substantial value of 1.35 $\mu_B$, reached in 50 kOe. In contrast, the magnetization measured at *T* = 10 K shows a linear field dependence up to the maximum field available, thus corroborating the paramagnetic state.

Figure 4 presents the temperature dependence of the electrical resistivity, *ρ(T)*, of platelet-like $CePtIn_4$ with electrical current flowing in the crystallographic *a-c* plane. Clearly, the compound is a metallic conductor in the entire temperature range investigated. At the onset of the AFM ordering, a sharp drop in *ρ(T)* is seen (see the inset to Figure 4), which occurs due to reduction in spin disorder scattering. Interestingly, no similar feature was seen before neither for the NE-type crystals nor the Zn-stabilized crystals of $CePtIn_4$ at their respective AFM phase transitions [15, 25]. The Neel temperature, defined as a maximum in the temperature derivative of the resistivity, $\frac{d\rho}{dT}(T)$, amounts to 2.3 K, in very good agreement with the magnetic data.

Overall, the electrical behavior of PL-type $CePtIn_4$ is similar to that found before for the needle-like crystals [15]. However, one should notice some important differences in the data measured for the two representatives. While the residual resistivity, $\rho_0$, of the NE-type crystal was as small as 0.1(5) μΩcm [15], it is about 8(2) μΩcm in the present case. Similarly, the resistivity observed at room temperature for the NE-type crystal was only 10 μΩcm, while it amounts to about 105 μΩcm for the PL-type one. In consequence, the residual resistivity ratio of the latter crystal RRR = 13 is significantly smaller than RRR = 100 reported before [15]. These striking differences might be partly attributed to strong anisotropy in the electrical behavior of single-crystalline $CePtIn_4$, recalling that the experiment in Ref. [15] was carried out along the crystallographic *a* axis, whereas in the present study the resistivity was probed with current flowing in the *a-c* plane. However, probably, the main factor governing the excellent electrical conductivity in the NE-type crystals of $CePtIn_4$ was ultra-high metallurgical quality of the whisker-like specimen measured.

To elucidate bulk and intrinsic nature of the AFM ordering in the PL-type crystals of $CePtIn_4$, heat capacity measurements, *C*(*T*), were performed down to 0.4 K. As shown in Figure 5, at room temperature, the specific heat is close to the Dulong-Petit limit $3nR$ = 149.6 J/(mol K), where *n* represents the number of atoms per formula units (here *n* = 6), and *R* = 8.314 J/(mol K) is the universal gas constant. With decreasing temperature, the specific heat decreases smoothly down to $T_N$ = 2.3 K, at which the AFM order sets in. A sharp λ-shaped anomaly at $T_N$ with a sizeable heat capacity jump $\Delta C$ = 7.6(6) J/(mol K) (see the inset to Figure 5) manifests the bulk nature of the phase transition and its second-order character. Similar results were obtained before for the needle- like crystals of $CePtIn_4$ [15]. However, at odds with the previous data, *C/T*(*T*) of the PL-type crystals exhibits an additional broad hump like feature near 1 K (marked by vertical arrow in the inset to Figure 5), i.e. in the AFM state. This highly puzzling feature is presently a subject of our supplementary investigations, the results of which will be reported elsewhere. From the low-temperature specific heat data, the magnetic entropy, *S*(*T*), was calculated, as illustrated in the inset to Figure 5. It appeared that the entropy released by $T_N$ amounts to *S* = 4.4 J/(mol K), which is only 76% of the value *R*ln2 = 5.76 J/(mol K) expected for magnetic ordering with a doublet ground state. This finding is also a subject of our forth-going study on the low-temperature behavior of PL-type $CePtIn_4$.

### 3.3. Electronic structure calculations

The configuration with the lowest energy among the three considered antiferromagnetic models is shown in Figure 6 together with the resultant densities of electronic states (DOS) calculated within GGA with orbital polarization corrections. The valence band starts about -10 eV below the Fermi level. The bottom of the band is formed mainly by *s*-type electrons and the central part consists of *d*-type electrons, mainly from Pt. The Ce atoms contributions cross the Fermi level and remain largely unoccupied. They indicate antiferromagnetic spin polarization and spin-orbit splitting and consist mainly of 4*f* states. The main contribution to the DOS at the Fermi level is provided by the Ce orbitals, about 87% of the total value of 22.4 states/eV/f.u., which leads to the Sommerfeld coefficient $\gamma$ = 58 mJ/(mol $K^2$). The spin and orbital magnetic moments on the Ce atoms are antiparallel and equal to -0.98 and 1.94 $\mu_B$/atom, respectively, leading to the total magnetic moments equal to ±0.96 $\mu_B$/atom. The orbital moment is significantly increased by the application of the orbital polarization corrections to the Ce 4*f* orbitals. The magnetic moments induced on the other atomic sites in the unit cell of $CePtIn_4$ are smaller than 0.01 $\mu_B$.

## 4. Conclusion

In summary, we have grown platelet-like single crystals of $CePtIn_4$, which were established within the experimental accuracy to possess the same chemical composition and the same orthorhombic crystal structure as the needle-like crystals of the compound, which were investigated by us before [15]. Comparing the crystal data of the PL-type and NE-type crystals one finds out that the crystallographic unit cell of the former variant is about 1.6 % larger than that of the latter one.

Similar to NE-type $CePtIn_4$, its PL-type variant also shows Curie-Weiss paramagnetism due to the presence of fairly stable $Ce^{3+}$ ions and orders antiferromagnetically at low temperatures. However, the Néel temperature of the platelet-like crystals ($T_N$ = 2.30(2) K) is distinctly higher than that of the needles ($T_N$ = 1.65(1) K). We anticipate that this difference may reflect the effect of chemical pressure, and can be rationalized in terms of the Doniach phase diagram [26], invoking dissimilar hybridization strength between 4*f* electronic states of cerium and electronic orbitals of neighboring atoms, mostly *p*-states of indium. Further experimental studies of single-crystalline $CePtIn_4$ with both platelet-like and needle-like morphologies as well as more detailed electronic band structure calculations for both crystalline variants of this compound are presently underway to shed more light on the actual microscopic origin of the discrepancies in their low-temperature physical properties.

**Acknowledgement**

The work was supported by the National Science Centre (Poland) under research grant No. 2015/19/B/ST3/03158. MW acknowledges the financial support of the National Science Centre Poland under the decision DEC- 2018/30/E/ST3/00267. Part of the computations were performed on the resources provided by the Poznań Supercomputing and Networking Center (PSNC).

**Supplementary information**

Further details on the present structural study can be obtained from the Fachinformationzentrum Karlsruhe, D-76344 Eggenstein-Leopoldshafen, Germany (http://www.fiz-karlsruhe.de/), quoting the depository number CCDC 1907453.

## References

[1] G.R. Stewart, Rev. Mod. Phys. **56**, 755 (1984).

[2] G.R. Stewart, Rev. Mod. Phys. **73**, 797 (2001).

[3] F. Steglich, J. Phys. Soc. Jpn **74**, 167 (2005).

[4] C. Petrovic, P.G. Pagliuso, M.F. Hundley, R. Movshovich, J.L. Sarrao, J.D. Thompson, Z. Fisk and P. Monthoux. J. Phys.: Condens. Matter **13**, L337 (2001).

[5] D. Kaczorowski, A.P. Pikul, D. Gnida and V.H. Tran, Phys. Rev. Lett. **103**, 027003 (2009); ibid. **104**, 059702 (2010).

[6] D. Kaczorowski, D. Gnida, A.P. Pikul and V.H. Tran, Solid State Commun. **150**, 411(2010).

[7] M.Werwiński, A.Szajek, A.Ślebarski, and D.Kaczorowski, J. Alloys Compd. **647**, 605 (2015).

[8] J. Prokleška, M. Kratochvílová, K. Uhlířová, V. Sechovský and J. Custers, Phys. Rev. B **92**, 161114 (2015).

[9] J. Custers, M. Diviš and M. Kratochvílová, J. Phys.: Conf. Ser. **683**, 012005 (2016).

[10] D. Das, D. Gnida, L. Bochenek. A. Rudenko, M. Daszkiewicz and D. Kaczorowski, Sci. Rep. **8**, 16703 (2018).

[11] M. Kratochvílová, J. Prokleška, K. Uhlířová, V. Tkáč, M. Dušek, V. Sechovský and J. Custers, Sci. Rep. **5**, 15904 (2015).

[12] D. Das, D. Gnida and D. Kaczorowski, Phys. Rev. B **99**, 054425 (2019).

[13] M. Falkowski and A.M. Strydom, J. Alloys Compd. **613**, 204 (2014).

[14] D. Das and D. Kaczorowski, J. Magn. Magn. Mater. **471**, 315 (2019).

[15] J. Bławat, D. Gnida, M. Daszkiewicz, P. Wiśniewski and D. Kaczorowski, J. Alloys Compd. **724**, 581 (2017).

[16] P. Hohenberg, W. Kohn, Phys. Rev. **136,** B864 (1964).

[17] G. M. Sheldrick, Acta Crystallogr. A **71**, 3 (2015).

[18] Diamond 3.2k - Crystal and Molecular Structure Visualization, Crystal Impact - Dr. H. Putz & Dr. K. Brandenburg GbR, Kreuzherrenstr. 102, 53227 Bonn, Germany, http://www.crystalimpact.com/diamond.

[19] K. Momma, F. Izumi, J. Appl. Crystallogr. **41**, 653 (2008).

[20] K. Koepernik, H. Eschrig, Phys. Rev. B **59,** 1743 (1999).

[21] J.P. Perdew, K. Burke, M. Ernzerhof, Phys. Rev. Lett. **77,** 3865 (1996).

[22] P.E. Blöchl, O. Jepsen, O.K. Andersen, Phys. Rev. B **49**, 16223 (1994).

[23] M.S.S. Brooks, Physica B+C **130,** 6 (1985).

[24] O. Eriksson, M.S.S. Brooks, B. Johansson, Phys. Rev. B **41,** 7311(1990).

[25] E.M. Carnicom, T. Klimczuk, F. von Rohr, M.J. Winiarski, T. Kong, K. Stolze, W. Xie, S.K. Kushwaha and R.J. Cava, J. Phys. Soc. Japan **86**, 084710 (2017).

[26] S. Doniach, *Physica B* **91**, 231 (1977).

Table 1. Crystal data and structure refinement details for PL-type $CePtIn_4$.

| | |
|---|---|
| *Crystal system* | orthorhombic |
| *Space group* | *Cmcm* |
| $a$, $b$, $c$ (Å) | 4.52175 (15), 16.7777 (5), 7.3903 (2) |
| $V$ (Å$^3$) | 560.66 (3) |
| $Z$ | 4 |
| $\mu$ (mm$^{-1}$) | 48.81 |
| Crystal size (mm) | 0.08 × 0.05 × 0.03 |
| *Data collection* | |
| $T_{min}$, $T_{max}$ | 0.122, 0.299 |
| No. of measured, independent and observed [$I > 2\sigma(I)$] reflections | 8428, 418, 415 |
| $R_{int}$ | 0.026 |
| $(\sin \theta/\lambda)_{max}$ (Å$^{-1}$) | 0.666 |
| *Refinement* | |
| $R[F^2 > 2\sigma(F^2)]$, $wR(F^2)$, $S$ | 0.010, 0.019, 1.31 |
| No. of reflections/par. | 418/24 |
| $\Delta\rho_{max}$, $\Delta\rho_{min}$ (e Å$^{-3}$) | 0.63, –0.78 |

Table 2. Atomic coordinates and equivalent isotropic displacement parameters ($Å^2$ x $10^3$) for the crystal structure of PL-type $CePtIn_4$. U(eq) is defined as one third of the trace of the orthogonalized $U^{ij}$ tensor.

| | X | y | z | U(eq) |
|---|---|---|---|---|
| Ce(1) | 0 | 0.1233(1) | 0.25 | 11(1) |
| Pt(1) | 0 | 0.2253(1) | 0.75 | 9(1) |
| In(1) | 0 | 0.5 | 0.5 | 14(1) |
| In(2) | 0 | 0.0669(1) | 0.75 | 13(1) |
| In(3) | 0 | 0.3169(1) | 0.4506(1) | 9(1) |

Table 3. Anisotropic displacement parameters ($Å^2$ x $10^3$) of the atoms in the crystal structure of PL-type $CePtIn_4$. The anisotropic displacement factor exponent takes the form: $-2\pi^2[ h^2 a^{*2}U^{11} + ... + 2 h k a^* b^* U^{12} ]$.

| | $U^{11}$ | $U^{22}$ | $U^{33}$ | $U^{23}$ | $U^{13}$ | $U^{12}$ |
|---|---|---|---|---|---|---|
| Ce(1) | 8(1) | 9(1) | 15(1) | 0 | 0 | 0 |
| Pt(1) | 7(1) | 9(1) | 10(1) | 0 | 0 | 0 |
| In(1) | 13(1) | 10(1) | 20(1) | 0(1) | 0 | 0 |
| In(2) | 12(1) | 9(1) | 18(1) | 0 | 0 | 0 |
| In(3) | 8(1) | 11(1) | 8(1) | 0(1) | 0 | 0 |

Table 4. Interatomic distances [Å] in the unit cell of PL-type $CePtIn_4$.

| | | |
|---|---|---|
| Ce(1) | 1 In(2) | 3.1912(5) |
| | 4 In(3) | 3.3188(2) |
| | 2 Pt(1) | 3.4013(3) |
| | 2 In(3) | 3.5713(4) |
| | 4 In(1) | 3.5780(2) |
| | 2 In(2) | 3.8141(1) |
| Pt(1) | 1 In(2) | 2.6564(5) |
| | 2 In(3) | 2.6944(3) |
| | 4 In(3) | 2.79473(19) |
| | 2 Ce(1) | 3.4013(3) |
| In(1) | 2 In(3) | 3.0933(3) |
| | 4 In(2) | 3.12833(17) |
| | 4 Ce(1) | 3.5780(2) |
| In(2) | 1 Pt(1) | 2.6564(5) |
| | 4 In(1) | 3.12834(17) |
| | 1 Ce(1) | 3.1913(5) |
| | 4 In(3) | 3.3326(4) |
| | 2 Ce(1) | 3.8141(1) |
| In(3) | 1 Pt(1) | 2.6944(3) |
| | 2 Pt(1) | 2.79474(19) |
| | 1 In(3) | 2.9651(6) |
| | 2 In(3) | 3.2690(4) |
| | 2 Ce(1) | 3.3188(2) |
| | 2 In(2) | 3.3327(4) |
| | 1 Ce(1) | 3.5713(4) |

**Figure captions**

Figure 1. Photograph of as-grown NE-type and PL-type single crystals of $CePtIn_4$ taken under optical microscope on a millimeter grid.

Figure 2. (a) Schematic view of the crystallographic unit cell of PL-type $CePtIn_4$ projected on the *b*-*c* plane. (b) Coordination spheres of the Ce and Pt atoms.

Figure 3. Temperature dependence of the reciprocal magnetic susceptibility of PL-type $CePtIn_4$ measured in a magnetic field of 3 kOe applied along the crystallographic *b* axis. The solid red line represents least-squares fit of Curie-Weiss law to the experimental data above 50 K. Upper inset: low-temperature variation of the magnetic susceptibility taken in $H$ = 1 kOe parallel to the *b* axis. Lower inset: field dependencies of the magnetization measured at $T$ = 1.8 K (ordered state) and 10 K (paramagnetic state) in magnetic field along the *b* axis.

Figure 4. Temperature dependence of the electrical resistivity of PL-type $CePtIn_4$ measured with electric current flowing within the crystallographic *a*-*c* plane. Inset: low-temperature resistivity data and temperature derivative of the electrical resistivity in the region of antiferromagnetic phase transition.

Figure 5. Temperature dependence of the specific heat of PL-type $CePtIn_4$. Inset: low-temperature data plotted as specific heat over temperature ratio and magnetic entropy in the region of antiferromagnetic phase transition.

Figure 6. The spin-projected total (upper panel) and partial (lower panel) densities of states for PL-type $CePtIn_4$. Inset: the antiferromagnetic arrangement of the magnetic moments in the ground state antiferromagnetic configuration considered. In the presented unit cell, Ce1 and Ce2 atoms are depicted by red and blue spheres, respectively, Pt atoms by green spheres, and In atoms by yellow spheres.

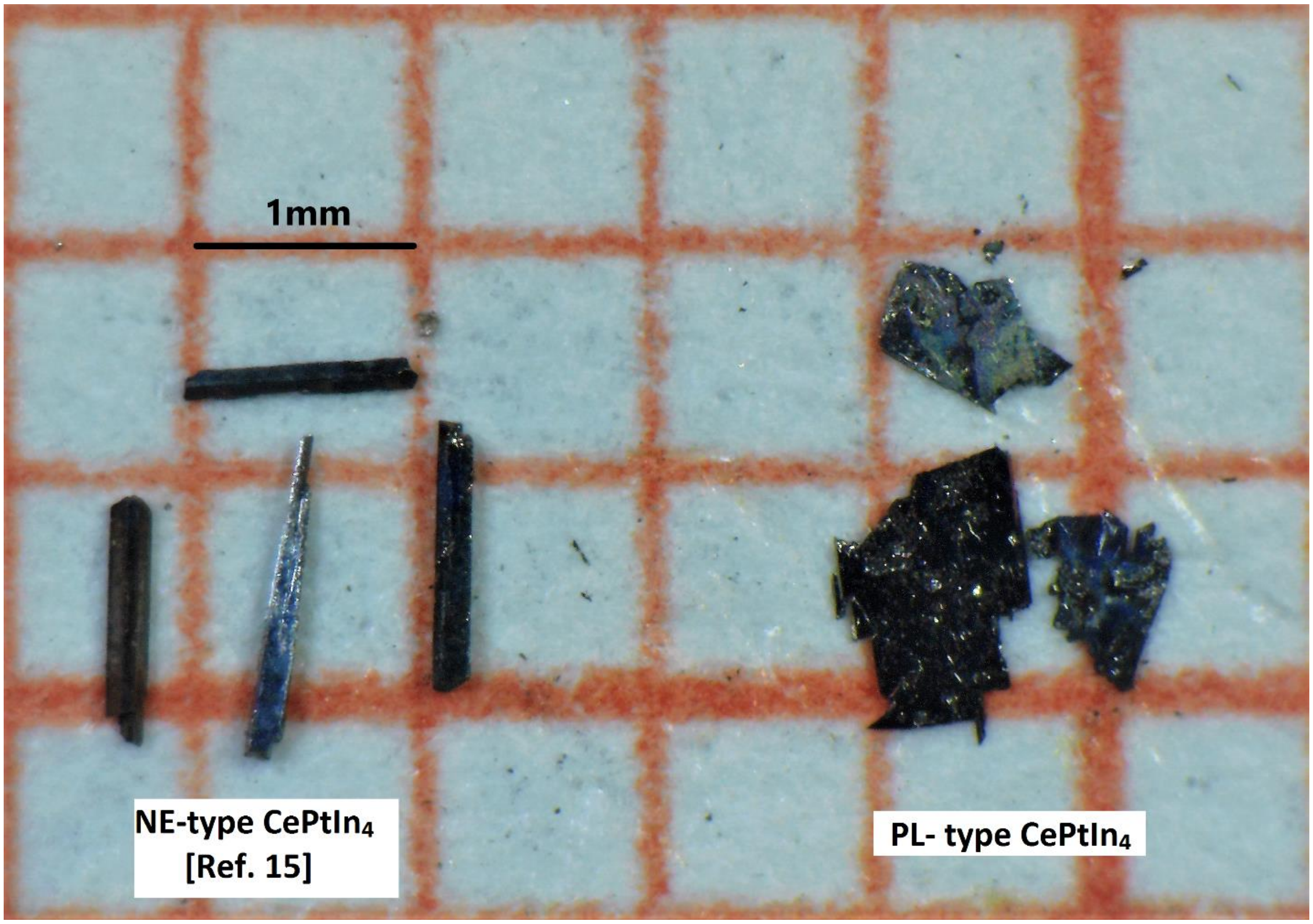

Figure 1.

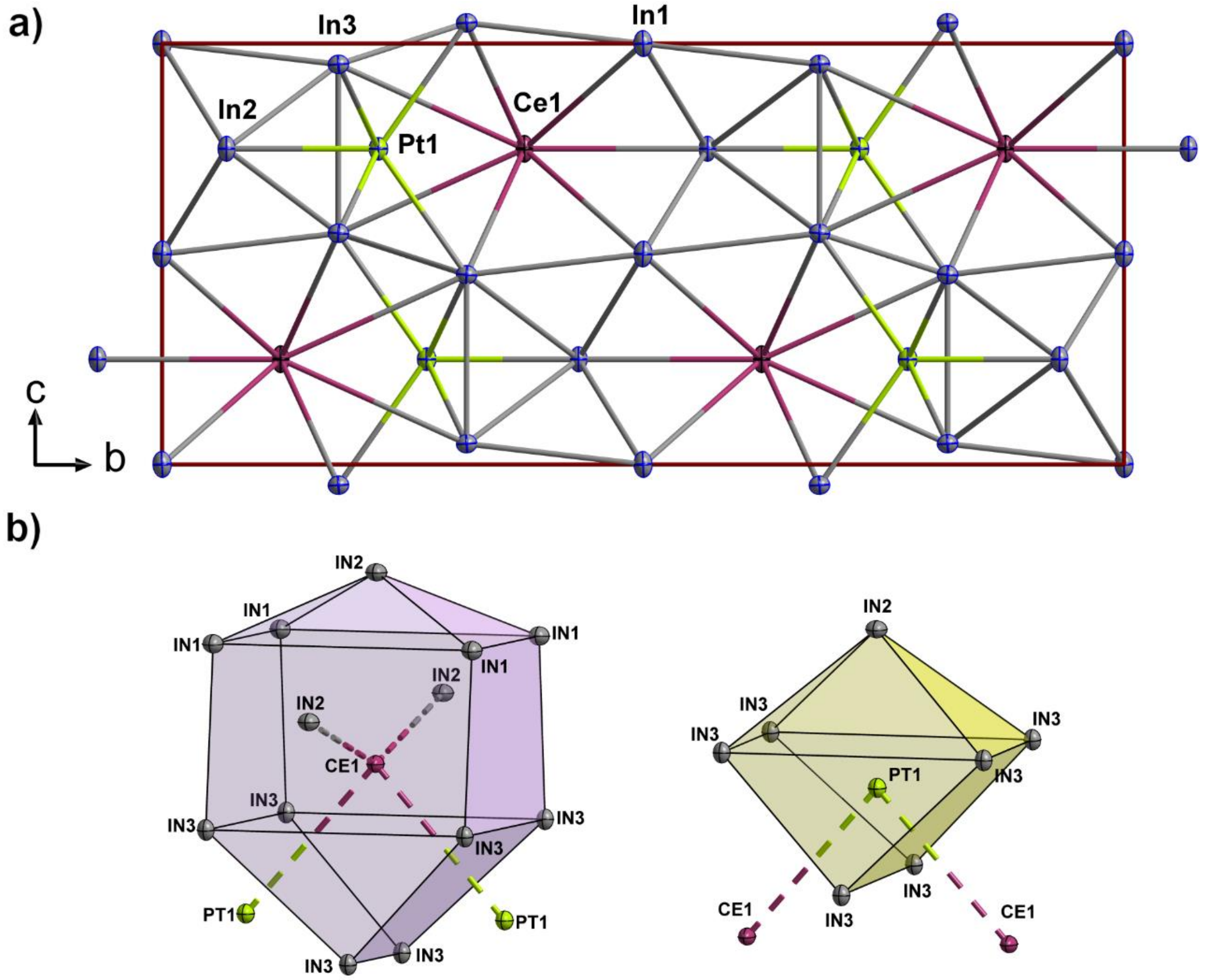

Figure 2.

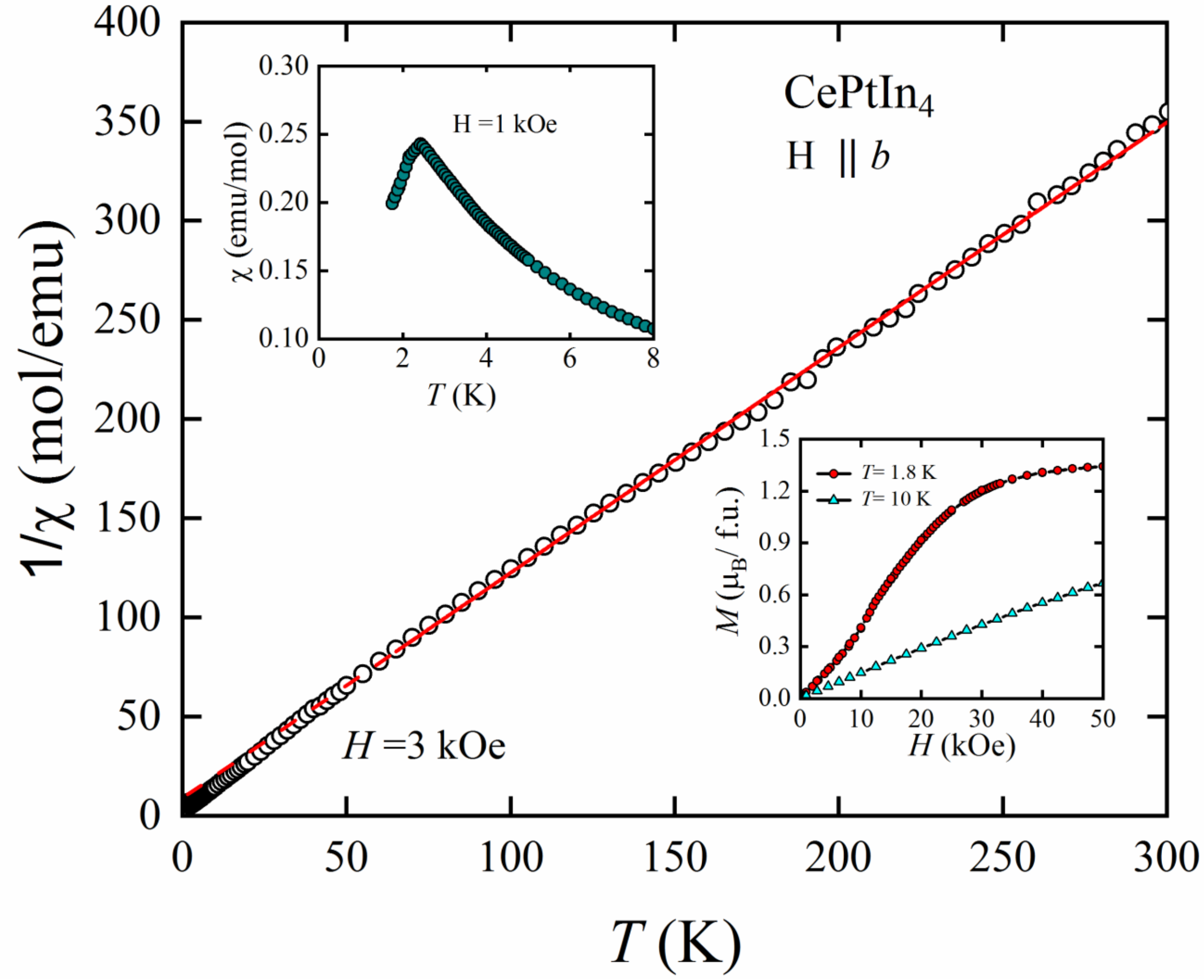

Figure 3.

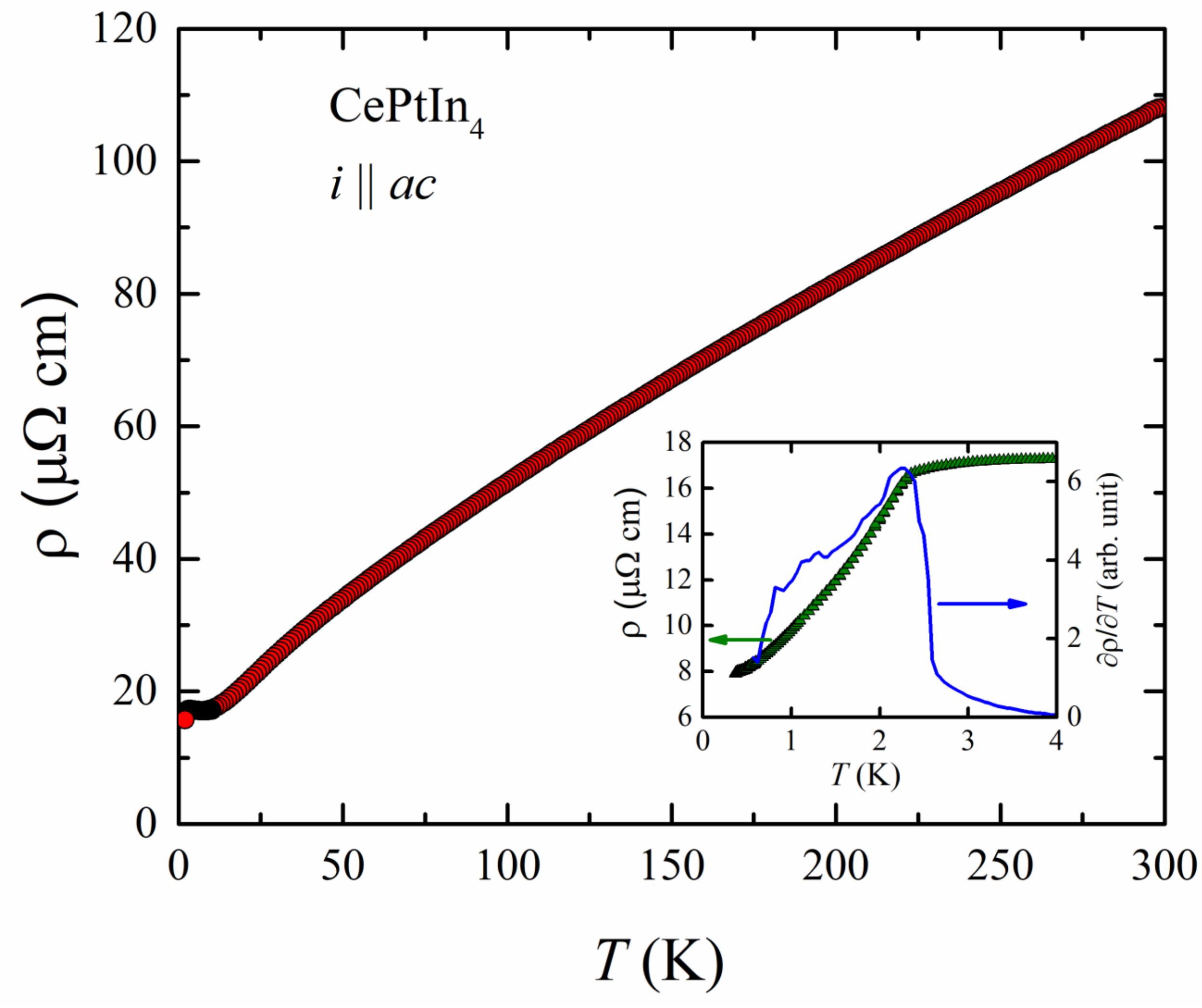

Figure 4.

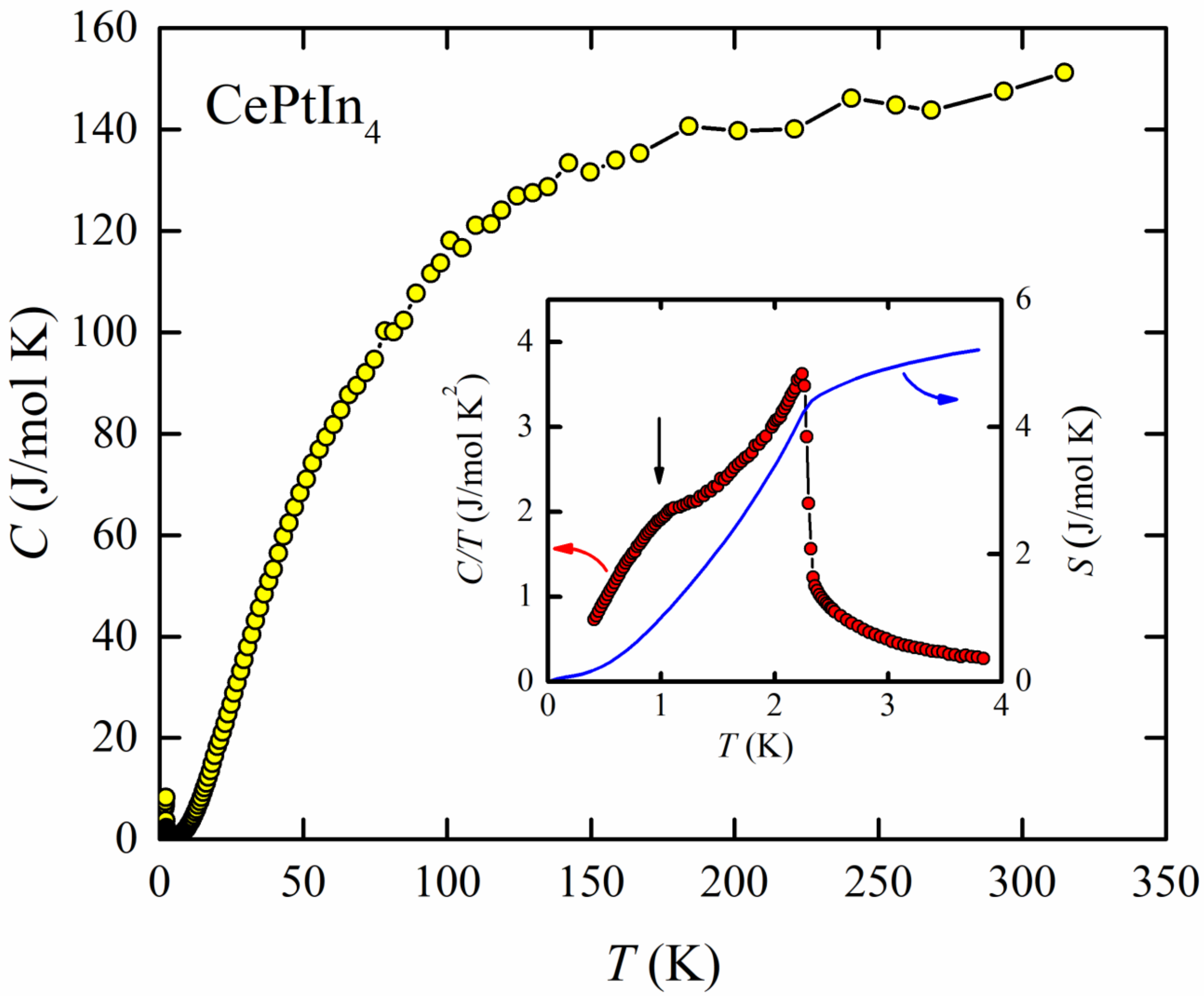

Figure 5.

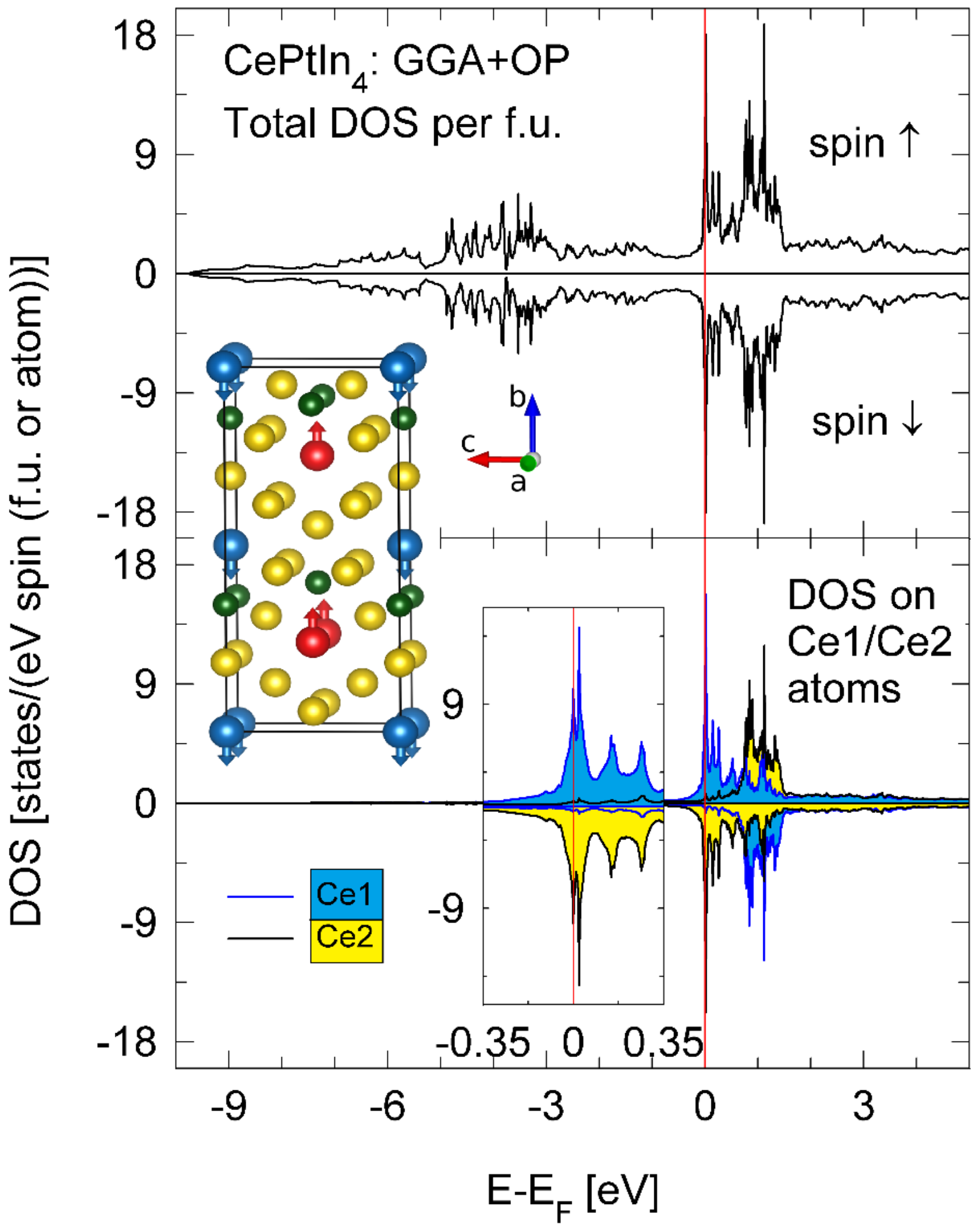

Figure 6.